\documentclass[aps,pre,reprint,superscriptaddress,longbibliography]{revtex4-1}
\usepackage{amsmath,amssymb,graphicx,bm}
\begin{document}

\title{Spin-orbit interactions in optically active materials}

\author{Chandroth P. Jisha}
\affiliation{ Centro de F\'{\i}sica do Porto, Faculdade de Ci\^encias, Universidade do Porto, R. Campo Alegre 687, Porto 4169-007, Portugal.}

\author{Alessandro Alberucci}

\affiliation{Optics Laboratory, Tampere University of Technology, FI-33101 Tampere, Finland}
\email{alessandro.alberucci@gmail.com}

%\ociscodes{(160.1190)   Anisotropic optical materials; (260.1440) Birefringence; (260.2110)   Electromagnetic optics; (260.2710)   Inhomogeneous optical media }

%\doi{\url{http://dx.doi.org/10.1364/ao.XX.XXXXXX}}

\begin{abstract}
We investigate the inherent influence of light polarization on the intensity distribution in anisotropic media undergoing a local inhomogeneous rotation of the principal axes. Whereas in general such configuration implies a complicated interaction between geometric and dynamic phase, we show that, in a medium showing an inhomogeneous circular birefringence, the geometric phase vanishes. Due to the spin-orbit interaction, the two circular polarizations perceive reversed spatial distribution of the dynamic phase. Based upon this effect, polarization-selective lens, waveguides and beam deflectors are proposed.
\end{abstract}

\date{\today}

\maketitle
%\thispagestyle{fancy}

%\ifthenelse{\boolean{shortarticle}}{\ifthenelse{\boolean{singlecolumn}}{\abscontentformatted}{\abscontent}}{}

%\section{Introduction}

The simplest approaches in wave optics model light as a scalar wave, an approximation valid only for paraxial beams propagating in isotropic homogeneous media \cite{Lax:1975}. The electromagnetic nature of light implies that photons have spin, appearing in the Maxwell's equations as the field polarization. Nonetheless, in case of paraxial waves propagating in isotropic materials, the spatial degree of freedom (i.e., the field distribution) is independent of the polarization. This is not rigorously true due to the vectorial nature of the Maxwell's equation, leading to a so-called \textit{classical} entanglement \cite{Spreeuw:1998,Berg-Johansen:2015}. The discrepancies between the two approaches are significant, for example, in case of non-paraxial beams leading to polarization-dependent focusing \cite{Dorn:2003}, or for polarization-dependent trajectory of light in inhomogeneous materials, the so called optical Magnus effect, aka spin Hall effect (SHE) of light \cite{Bliokh:2004,Bliokh:2007,Bliokh:2008,Zhang:2016}. 
 
In this context, the investigation of spin-orbit interaction is rapidly becoming a central topic in optics \cite{Bliokh:2015}. Photons have been demonstrated to be a unique tool for the investigation of basic quantum field theory \cite{Neugenbauer:2015,Bliokh:2015_1}. On the other hand, spin-orbit interaction paves the way to a new family of ultra-thin photonic devices, including gratings \cite{Bomzon:2002}, lenses \cite{Lin:2014}, polarimeters \cite{Todorov:1992,BalthasarMueller:2016}, beam shaper \cite{Marrucci:2006,Arbabi:2015}, sensors \cite{Petersen:2014}, deflectors \cite{Li:2013} and so on. 

In essence, spin-orbit interactions in isotropic materials are connected with the fact that singular plane waves composing the beam are transverse electromagnetic in different frameworks. Practical examples include beams bended by gradient index \cite{Bliokh:2008}, interaction with interfaces \cite{Hosten:2008} and even free space \cite{Dorn:2003}. These effects can be interpreted in terms of a geometric phase called the Rytov-Vladimirskii-Berry phase. 

In anisotropic media another type of geometric phase arises, the Pancharatnam-Berry phase (PBP). PBP appears in the presence of a rotation of the beam polarization for a fixed wave-vector. Noteworthy, in anisotropic materials spin-orbit effects, such as birefringence or spatial walk-off, are observed even for plane waves due to the dependency on polarization of the light-matter interaction \cite{Saleh:2007}. Nonetheless, the presence of PBP in anisotropic media widens the spectrum of the observable spin-orbit effects \cite{Bomzon:2002,Marrucci:2006,Lin:2014}, including the nonlinear case \cite{Li:2015,Nookala:2016}. The PBP affects the wavefront, and thus light propagation, when the anisotropic medium is inhomogeneous and showing a point-wise rotation of the principal axes across the intensity profile. In this Letter we first analyze light propagation in inhomogeneously rotated materials considering the trade-off between the PBP and diffraction. We show that PBP vanishes if the material is optically active,
 the field propagation then being affected by dynamic phase alone. Unlike Refs.~\cite{Bliokh:2004,Bliokh:2007,Zhang:2016} where geometric optics is used, here the wave behavior of the electromagnetic radiation is accounted for. Differently from the case of linear birefringence \cite{Alberucci:2016}, we demonstrate that in optically active media the two circular polarizations see a photonic potential reversed in sign, yielding polarization-dependent guiding and SHE. \\
Let us consider the monochromatic propagation of an electromagnetic field [amplitude $\propto \exp{\left(-i\omega t\right)}$] in an inhomogeneous anisotropic material. The average wavevector is parallel to the axis $z$.  We consider a non-magnetic material ($\bm{\mu}=\mu_0 \bm{I}$). 
Supposing that the medium permittivity does not vary along $z$, the electric field in the paraxial limit obeys
\begin{align}
 \nabla^2 \left( \begin{array}{c}
                                             E_x \\ E_y
                                            \end{array}\right)  + k_0^2 \left( \begin{array}{cc}
                                             \epsilon_{xx}(x,y) & \epsilon_{xy}(x,y) \\ 
																						\epsilon_{yx}(x,y) & \epsilon_{yy}(x,y)
                                            \end{array}\right) \cdot \left( \begin{array}{c}
                                             E_x \\ E_y
                                            \end{array}\right) =0,
																						 \label{eq:Maxwell}
\end{align}
where the anisotropy in the diffraction coefficients has been neglected and $k_0$ is the vacuum wavenumber. The paraxial hypothesis allows us to neglect the longitudinal field in \eqref{eq:Maxwell}. Finally, in writing \eqref{eq:Maxwell} we have assumed that the input beam, polarized on the plane $xy$, does not excite any appreciable polarization parallel to $\hat{z}$, that is, the linear walk-off vanishes. \\
Generally speaking, the two-dimensional tensor $\bm{\epsilon}_{2D}=(\epsilon_{xx},\epsilon_{xy};\epsilon_{yx},\epsilon_{yy})$ will change on the transverse plane $xy$ either due to a local change in the values of its eigenvalues and/or due to a local rotation of the principal axes. Variation of eigenvalues is associated with changes in the local refractive indices. Principal axes distribution, in our case, can be modeled like a rotation by an angle $\theta(x,y)$ around the propagation direction $z$, $\bm{R}(\theta)=\left(\cos\theta,\sin\theta;-\sin\theta,\cos\theta \right)$. Thus, for a generic vector $\bm{v}$ it is $(v_x^\prime;v_y^\prime)=\bm{R}(\theta)\cdot (v_x;v_y)$ where the subscript $^\prime$ indicates the framework of the local principal axes. To analyze the effect of the local rotation $\bm{R}(\theta)$, let us introduce the field in the rotated framework $(E_x^\prime;E_y^\prime)=\bm{R}(\theta)\cdot (E_x;E_y)$. After setting $ \bm{E}^\prime=E_x^\prime \hat{x} + E_y^\prime \hat{y}$, \eqref{eq:Maxwell}
 provides \cite{Alberucci:2016}
\begin{eqnarray}
 \label{eq:Maxwell_rotated}
 &\left(\frac{\partial^2}{\partial z^2} +\nabla^2_{xy} \right) \bm{E}^\prime + \bm{R}\cdot \nabla^2_{xy}\bm{R}^{-1}\cdot  \bm{E}^\prime + \nonumber \\
																							 & 2\bm{R}\cdot \left( \frac{\partial \bm{R}^{-1}}{\partial x} \cdot \frac{\partial }{\partial x} + 
																								\frac{\partial \bm{R}^{-1}}{\partial y} \cdot \frac{\partial }{\partial y}\right)   \bm{E}^\prime
																							+ k_0^2 \bm{\epsilon}^\prime_{2D} \cdot  \bm{E}^\prime =0.
\end{eqnarray} 
A straightforward computation yields
\begin{eqnarray}
 \label{eq:Maxwell_rotated_theta}
 &\left(\frac{\partial^2}{\partial z^2} +\nabla^2_{xy} \right) \bm{E}^\prime - \left[\left( \frac{\partial \theta}{\partial x}\right)^2 +   \left( \frac{\partial \theta}{\partial y}\right)^2 \right]      \bm{I}\cdot \bm{E}^\prime \nonumber \\  &- 	i\left(\frac{\partial^2 \theta}{\partial x^2} + \frac{\partial^2 \theta}{\partial y^2} \right) \bm{s}_2\cdot
																							   \bm{E}^\prime \nonumber \\	 & -2i \left(\frac{\partial \theta}{\partial x}  \frac{\partial}{\partial x} +
																						\frac{\partial \theta}{\partial y}  \frac{\partial}{\partial y} \right) \bm{s}_2 \cdot \bm{E}^\prime
																							+ k_0^2 \bm{\epsilon}^\prime_{2D} \cdot \bm{E}^\prime =0,
\end{eqnarray}
where $\bm{s}_j\ (j=1,2,3)$ represent the Pauli matrices. \\
Let us now discuss how the solutions of \eqref{eq:Maxwell_rotated_theta} depend on the features of $\bm{\epsilon}_{2D}^\prime$. From  a general point of view, transparent anisotropic materials  can be classified into two categories depending on the polarization of the plane waves which are supported as eigenmodes by the medium: we talk of linear or of circular birefringence when the eigenmodes are linearly or circularly polarized waves, respectively \cite{Saleh:2007}. The last family can be easily generalized to the case of elliptic polarizations. Mathematically, the anisotropic material shows linear birefringence if the components of $\bm{\epsilon}^\prime_{2D}$ are symmetric, whereas circular birefringence is related with an anti-symmetric $\bm{\epsilon}^\prime_{2D}$. Media fulfilling the latter condition are called gyrotropic due to the appearance of optical activity, and includes both chiral and magneto-optical materials, where the optical activity is spontaneous or induced by an external magnetic 
field, respectively \cite{Condon:1937,Saleh:2007}. For a lossless medium, the components should be also complex conjugate for the tensors to be Hermitian, that is,  $\epsilon_{i,j} = \epsilon_{j,i}^\ast$.\\
In the presence of linear birefringence, $\bm{\epsilon}_{2D}^\prime$ is diagonal with two distinct eigenvalues, let us call them $\epsilon_1$ and $\epsilon_2$. Then we can always set $\bm{\epsilon}^\prime_{2D}=\frac{\epsilon_1+\epsilon_2}{2}\bm{I}+\frac{\epsilon_a}{2} \bm{s}_3$, where $\epsilon_a=\epsilon_2-\epsilon_1$ is the optical anisotropy. Substituting back into \eqref{eq:Maxwell_rotated_theta}, we find terms proportional to two different Pauli matrices, thus \eqref{eq:Maxwell_rotated_theta} cannot be diagonalized in any basis \cite{Weinberg:2012}. This means that the polarization of the field will always vary while evolving along $z$, in turn leading to the occurrence of a transversely-varying PBP, whenever $\theta$ is not uniform. In accordance with the former statement, in the case of linear birefringence it is well known that the wavefront is strongly affected by the PBP, both for short \cite{Bomzon:2002,Marrucci:2006,Li:2013} or long propagation \cite{Slussarenko:2016,Alberucci:2016} with respect 
to the Rayleigh distance. \\
Light propagation changes drastically in the presence of circular birefringence, that is, $\bm{\epsilon}^\prime_{2D}=(\epsilon_1,-i\gamma;i\gamma,\epsilon_2)$, where $\gamma$ is the modulus of the gyration vector. As a matter of fact, if $\epsilon_1=\epsilon_2$ then $\bm{\epsilon}^\prime_{2D}$ is proportional to $\bm{s}_2$, and \eqref{eq:Maxwell_rotated_theta} can be diagonalized in the circular basis composed by left circular polarization (LCP) $\hat{L}=\left(\hat{x}- i\hat{y}\right)/\sqrt{2}$ and right circular polarization (RCP) $\hat{R}=\left(\hat{x} + i\hat{y}\right)/\sqrt{2}$. Complex unit vectors $\hat{L}$ and $\hat{R}$ can also be rewritten as $\left(\hat{x}+i\sigma \hat{y}\right)/\sqrt{2}$, where $\sigma$ is the photon helicity, with $\sigma_L=-1$ and $\sigma_R=1$. Hence, no exchange of power between the two circular polarizations takes place versus $z$. This behavior agrees with physical intuition: a rotation around the propagation distance does not change the polarization state for circular 
polarizations, the latter being eigenvectors of the rotation 
matrix $\bm{R}(\theta)$.\\
Above discussion holds valid even when $\gamma$ varies across $xy$ (variations must be small on the wavelength scale, otherwise \eqref{eq:Maxwell} is no more valid). Formally, let us introduce the slowly varying envelopes $\psi_l=E_l \exp{(-ik_0n_0z)}\ (l=x,y)$, where $n_0=\sqrt{\epsilon_1}=\sqrt{\epsilon_2}$. Then \eqref{eq:Maxwell_rotated_theta} can be diagonalized using the space-independent transformation between linear and circular polarization basis,
\begin{equation}
  \left( \begin{array}{c} \psi_L \\ \psi_R
  \end{array} \right) = \frac{1}{\sqrt{2}} \left(\begin{array}{cc} 1 & i \\ 1 & -i
  \end{array} \right) \cdot \left(\begin{array}{c} \psi_x \\ \psi_y
  \end{array} \right). \label{eq:transformation_linear_circular}
\end{equation}
Substitution of \eqref{eq:transformation_linear_circular} to \eqref{eq:Maxwell_rotated_theta} yields the two uncoupled equations
\begin{eqnarray}
 \label{eq:NLSE_circular_L}
 &2ik_0n_0 \frac{\partial \psi_L}{\partial z}  +\nabla^2_{xy} \psi_L - \left[\left( \frac{\partial \theta}{\partial x}\right)^2 +   \left( \frac{\partial \theta}{\partial y}\right)^2 \right]  \psi_L   + 	i\left(\frac{\partial^2 \theta}{\partial x^2} + \frac{\partial^2 \theta}{\partial y^2} \right) \psi_L \nonumber \\	 & +2i \left(\frac{\partial \theta}{\partial x}  \frac{\partial}{\partial x} + \frac{\partial \theta}{\partial y}  \frac{\partial}{\partial y} \right) \psi_L	- k_0^2 \gamma(x,y) \psi_L =0,  \\
&2ik_0n_0 \frac{\partial \psi_R}{\partial z}  +\nabla^2_{xy} \psi_R - \left[\left( \frac{\partial \theta}{\partial x}\right)^2 +   \left( \frac{\partial \theta}{\partial y}\right)^2 \right]  \psi_R   - 	i\left(\frac{\partial^2 \theta}{\partial x^2} + \frac{\partial^2 \theta}{\partial y^2} \right) \psi_R \nonumber \\	 & -2i \left(\frac{\partial \theta}{\partial x}  \frac{\partial}{\partial x} + \frac{\partial \theta}{\partial y}  \frac{\partial}{\partial y} \right) \psi_R	+ k_0^2  \gamma(x,y) \psi_R =0.  \label{eq:NLSE_circular_R}
\end{eqnarray}
Eventually, application of the polarization-dependent gauge transformation $A_p=\psi_p\exp{[-i\sigma_p \theta(x,y)]}\ (p=L,R)$ transforms Eqs.~(\ref{eq:NLSE_circular_L}-\ref{eq:NLSE_circular_R}) into two paraxial Helmholtz equations $2ik_0n_0\frac{\partial A_p}{\partial z}+\nabla^2_{xy} A_p + k_0^2 \sigma_p \gamma(x,y)A_p=0\ (p=L,R)$. There is no exchange of power between the LCP and the RCP, meaning no contribution from the PBP: only a polarization-dependent and inhomogeneous dynamic phase $\sigma_p \gamma(x,y)$ is acting on the electromagnetic field. Summarizing, light propagates as if subject to a polarization-dependent distribution of refractive index given by
\begin{align} \label{eq:index_RCP}
  &RCP:\ n_\mathrm{eff}(x,y) = \sqrt{n_0^2+\gamma(x,y)},\\
	&LCP:\ n_\mathrm{eff}(x,y) = \sqrt{n_0^2-\gamma(x,y)}. \label{eq:index_LCP}
\end{align}

We tested the theoretical predictions by means of in-house BPM (Beam Propagation Method) codes in (1+1)D and (2+1)D, simulating directly \eqref{eq:Maxwell} in the paraxial approximation (second derivative along $z$ is neglected) and in the presence of a transversely-varying circular birefringence. %in the linear polarization basis $(\hat{x},hat{y})$ .
Two types of inhomogeneity for $\gamma$ will be considered: with reference to the (1+1)D geometry, an even-symmetric (in two dimensions cylindrically symmetric) and an anti-symmetric profile (in two dimensions mirror symmetry will be imposed). According to Eqs.~(\ref{eq:index_RCP}-\ref{eq:index_LCP}), in the first case a focusing/defocusing behavior is expected according to the beam helicity, whereas in the second case a spin-dependent deflection should occur. \\
\begin{figure}
\centering
\includegraphics[width=1\columnwidth]{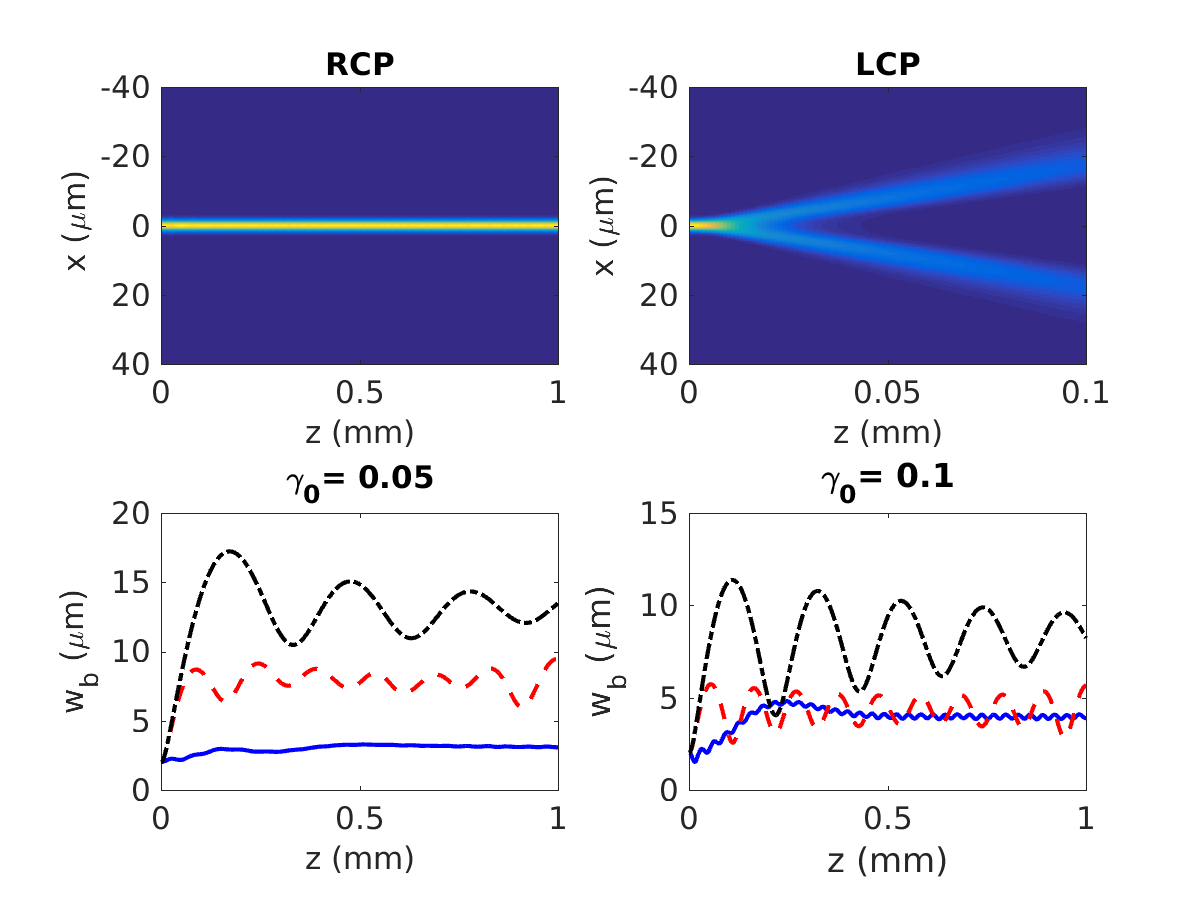}
\caption{Polarization-dependent confinement. Top row: light undergoes either guiding or anti-guiding behavior depending on the input polarization of the beam; for $w_g=2~\mu$m and $\gamma_0=0.05$  RCP (left) shows focusing and LCP (right) shows defocusing. Bottom row: beam width $w_b$ versus the propagation distance for varying strength of optical activity $\gamma_0$ and RCP input, as marked in each panel. Different curves correspond to different width $w_g$ of the optical activity profile: solid blue lines correspond to $w_g= 2\mu$m, dashed red lines correspond to $w_g= 10\mu$m and black dotted lines to $w_g= 20\mu$m. Input beam is Gaussian with a waist of $2~\mu$m placed in $z=0$.}
\label{fig:width1D}
\end{figure}
For simplicity, we begin with the one-dimensional case and later on extend the results to the two-dimensional case. 
For the even case, optical activity is supposed to be Gaussian, i.e., $\gamma = \gamma_0 \exp(-x^2/w_g^2)$, where $\gamma_0$ is the maximum of the optical activity and $w_g$ is the width of the transverse distribution. Hereafter the input beam is a fundamental Gaussian beam $\propto e^{-x^2/w_0^2}$ and a Laguerre-Gaussian (LG) beam $\propto r L_0^1(2r^2/w_0^2)	e^{-r^2/w_0^2} e^{i\phi}$, in the (1+1)D and (2+1)D cases respectively. We also take a wavelength of $1.064~\mu$m and $n_0=1$: the results can be easily generalized to other cases using standard normalizations. BPM results confirm the analytical results, i.e., the two polarizations evolve independently, seeing an effective index distribution provided by Eqs.~(\ref{eq:index_RCP}-\ref{eq:index_LCP}) (see top panels of Fig.~\ref{fig:width1D}). The bottom panels of Fig.~\ref{fig:width1D} depict the evolution of the beam width of the confined polarization for different values of the parameters, $\gamma_0$ and $w_g$, of the graded waveguide. The confining effect becomes stronger, e.g. the number of bounded modes raises up, either increasing $\gamma_0$ and/or the width of the optical activity profile $w_g$, as witnessed by the oscillations in the beam width. Power is almost completely coupled into the fundamental mode when $w_0=2~\mu$m, $w_g=2~\mu$m and $\gamma_0=0.05$  (first panel in Fig.~\ref{fig:width1D}).\\
\begin{figure}
\centering
\includegraphics[width=1\columnwidth]{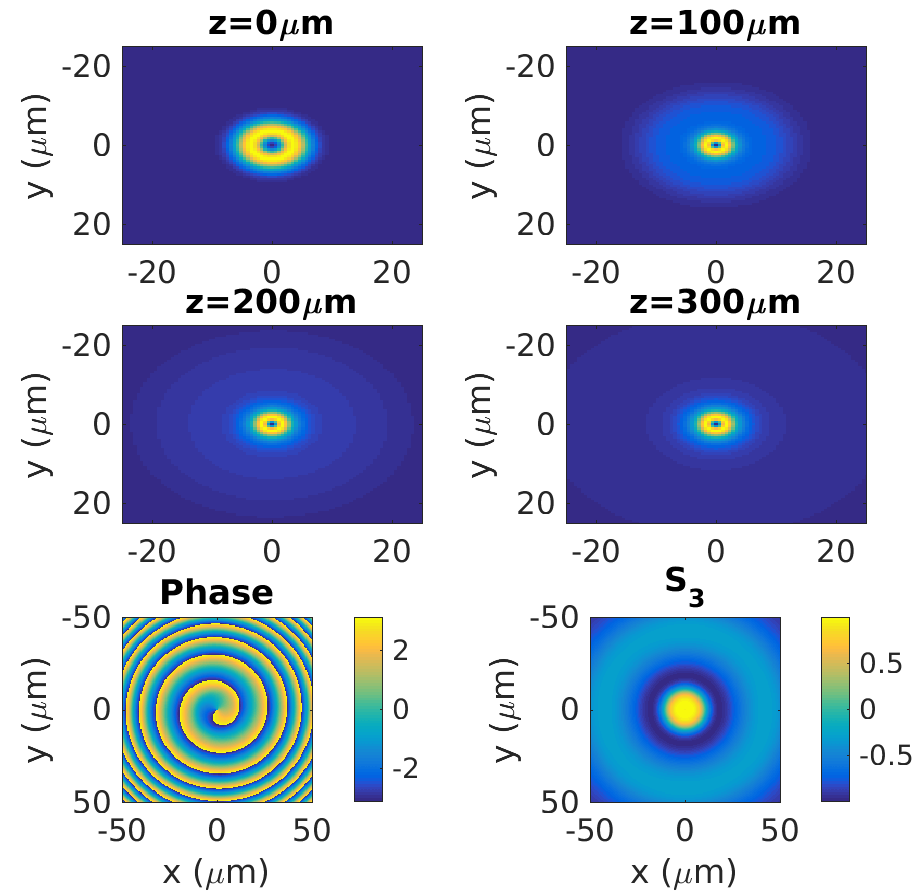}
\caption{Even symmetric $\gamma$: Profiles of the input LG beam with width $w_0=5~\mu$m at different propagation planes as marked (normalized with power at each plane). The confined beam corresponds to the RCP and the diffracting part corresponds to the LCP component of the beam. The last row shows the phase and the Stokes parameter $S_3$ for the beam at $z=300\mu$m. The parameters of the medium are $\gamma_0=0.05$ and $w_g=2~\mu$m.}
\label{fig:profile2D}
\end{figure}
Analogous behavior occurs in two-dimensional geometries. To achieve a complete description of light with one single simulation for each geometry, in (2+1)D  input beams are taken linearly polarized, that is, a superposition with the same weight of RCP and LCP waves. The RCP and LCP will be either focused and defocused, according to the sign of $\gamma_0$. To stress out the differences with respect to birefringent media \cite{Marrucci:2006}, we will consider beams carrying angular momentum. Accordingly, input beams are Laguerre-Gaussian beam carrying a unitary charge of orbital angular momentum (OAM), whereas for $\gamma$ a cylindrically-symmetric Gaussian profile is set. Results are plotted in Fig.~\ref{fig:profile2D}. The RCP component sees a focusing potential, coupling almost all its energy to guided modes of the waveguide. On the other hand, the LCP component perceives a repelling potential and spreads out faster than in a homogeneous medium. The transverse distribution of the beam phase, plotted in 
Fig.~\ref{fig:profile2D}, shows that the diffracting and the confined beam carry the same OAM of the input beam. In contrast, in an anisotropic medium with transversely varying optic axis (a so-called q-plate \cite{Marrucci:2006}), the spin angular momentum (SAM) and the OAM are inextricably connected due to the PBP. The polarization behavior is addressed plotting the $S_3$ component of the Stokes vector $\bm{S}$: for a purely RCP beam $S_3=1$, whereas for a LCP beam $S_3=-1$, with all the other Stokes components being equal for both the circular polarizations. Polarization distribution on the beam cross section $S_3$, plotted in Fig.~\ref{fig:profile2D}, shows that the confined part is RCP and the diffracting halo is LCP, fully confirming the predictions.\\
\begin{figure}
\centering
\includegraphics[width=1\columnwidth]{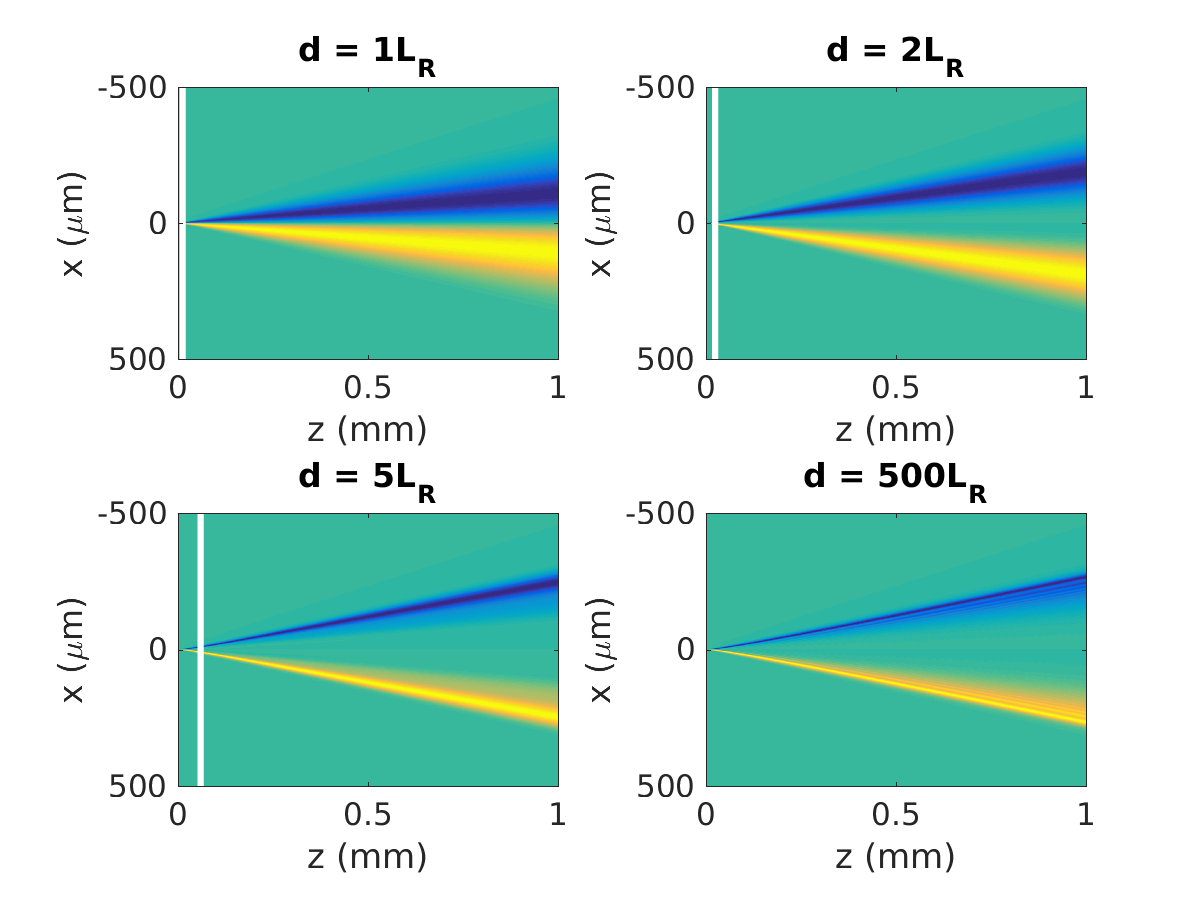}
\caption{Anti-symmetric $\gamma$ in (1+1)D: Evolution of the Stokes parameter $S_3$ showing the splitting of an input linearly polarized beam with $w_0 = 5~\mu$m. The beam in the upper and the lower half plane are LCP and RCP, respectively. Each panel corresponds to different longitudinal length $d$ (in the figure expressed in unit of the Rayleigh length $L_R$, equal to 11.8$~\mu$m in this case) of the medium, the latter extending from $z=0$ to the white lines (after free space is assumed). Here $\gamma_0=0.05$ and $w_g=2~\mu$m.}
\label{fig:soc1d}
\end{figure}
We change now the spatial symmetry of the optical activity, i.e., we consider an anti-symmetric profile for $\gamma$, $\gamma=\gamma_0 \tanh(x/w_g)$. Depending on the input polarization, the beam will get deflected either to the left or to the right, i.e., optical SHE \cite{Zhang:2016}. Figure~\ref{fig:soc1d} shows the evolution of the Stokes parameter $S_3$ in the (1+1)D case for a linearly polarized input. The RCP and LCP components separate spatially and are deflected symmetrically to either sides of the input direction. With increasing $\gamma_0$, the strength of deflection is increased (not shown here). Deflection angle also depends on the length of the optically active medium, see the different panels of Fig.~\ref{fig:soc1d}. Finally, for longer propagation lengths in the optically active medium, the beam begins to develop fringes due to self-interference effects (last panel). \\
\begin{figure}
\centering
\includegraphics[width=1\columnwidth]{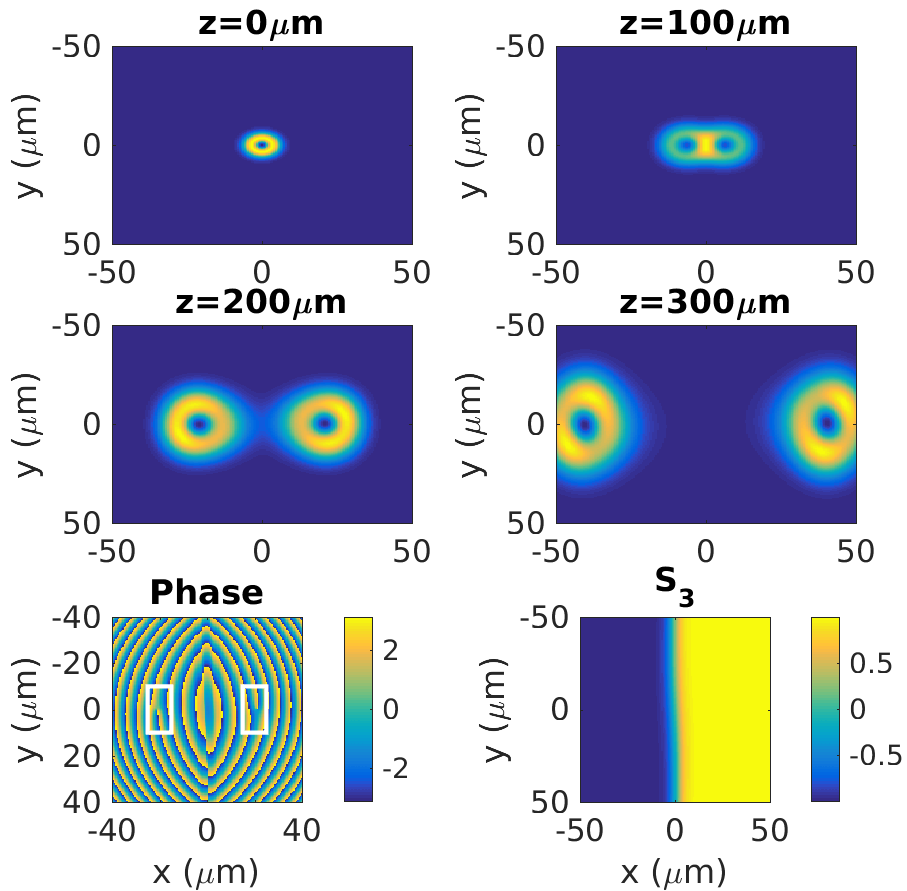}
\caption{Anti-symmetric $\gamma$ in (2+1)D: Profiles of the input LG beam with width $w_0=5~\mu$m at different propagation planes as marked. The right moving beam corresponds to the RCP and the left moving part corresponds to the LCP component of the beam. The last row shows the phase of the beam at $z=200~\mu$m and the Stokes parameter $S_3$ for the beam at $z=300~\mu$m. The white rectangles show the position of the two vortices, each of them carrying the original topological charge. The parameters of the medium are $\gamma_0=0.05$ and $w_g=20~\mu$m.}
\label{fig:profile2Dsoc}
\end{figure}
Next, we want to check the polarization-dependent deflection for beams carrying OAM. Input is a linearly polarized LG beam with unitary OAM charge. Like the SHE in the 1D case, the vortex beam splits into two specular halves, the LCP and the RCP. The two circular polarizations, deflected by the same angle in absolute value, broaden due to diffraction as well (Fig.~\ref{fig:profile2Dsoc}). The OAM of the two separated beams is same as that of the input, i.e., there is no conversion of SAM into OAM \cite{Marrucci:2006}.  \\ 
In conclusion, we demonstrated that, in locally twisted anisotropic materials, the Pancharatnam-Berry phase vanishes in the presence of circular birefringence. Unlike with linear birefringence, a stronger spin-orbit effect occurs due to the monotonic accumulation of the polarization-dependent effects in propagation. Although in natural media the smallness of optical activity hinders the observation of the proposed effects, man-made metamaterials with enhanced circular birefringence are ideal candidates to verify our theoretical results \cite{Plum:2007,Gansel:2009,Schaferling:2012,Kaschke:2015}, with possible extensions to the nonlinear case as well \cite{Ren:2012}. 

\textbf{Funding.} Funda\c{c}\~{a}o para a Ci\^{e}ncia e a Tecnologia, POPH-QREN and FSE (FCT, Portugal), fellowship SFRH/BPD/77524/2011. Academy of Finland FiDiPro no. 282858.

\end{document}